\documentclass[12pt]{article}
\usepackage{graphicx}
\usepackage{amssymb}

\def\bq{\begin{equation}}
\def\eq{\end{equation}}
\def\ba{\begin{array}}
\def\ea{\end{array}}
\def\e{\varepsilon}

\begin{document}
\begin{titlepage}

{\huge Stability calculations for the ytterbium-doped fiber laser passively
mode-locked through nonlinear polarization rotation}
\vspace{1.5cm}

M. Salhi, H. Leblond and F. Sanchez\\
{\it
Laboratoire POMA, UMR 6136, Universit\'e d'Angers
2 Bd Lavoisier, 49045 Angers Cedex, France}

M. Brunel and A. Hideur\\
{\it  Groupe d'Optique et d'Optronique, CORIA UMR 6614, Universit\'e de Rouen
Bd de L'Universit\'e BP 12, 76801 Saint-Etienne du Rouvray Cedex, France\vspace{1.5cm}}

{\bf Abstract}
\vspace{5mm}\\
We investigate theoretically a fiber laser passively mode-locked with nonlinear polarization rotation.
A unidirectional ring cavity is considered with a polarizer placed between two sets of a halfwave plate
and a quarterwave plate. A master equation is derived and the stability of the continuous and mode-locked
solutions is studied. In particular, the effect of the orientation of the four phase plates and of the
polarizer on the mode-locking regime is investigated.
\end{titlepage}

\section{Introduction}
Passively mode-locked fiber lasers are of great importance for
various applications involving optical telecommunications.
Different experimental methods have been used to achieve
mode-locking operation \cite{ref1}-\cite{ref11}. In this paper we
are interested in mode-locking through nonlinear polarization
rotation. This technique has been successfully used to obtain
short pulse generation in different rare-earth doped fiber lasers
\cite{ref3}-\cite{ref5},\cite{ref12}-\cite{ref15} and is
self-starting. The laser configuration is a unidirectional fiber
ring cavity containing a polarizer placed between two polarization
controllers. The polarization state evolves nonlinearly in the
fiber as a result of the optical Kerr effect. If the polarization
controllers are suitably oriented, the polarizer lets pass the
central intense part of a pulse while it blocks the low intensity
wings.

Different theoretical approaches have been developed to
describe the mode-locking properties of such laser. Haus {\it et al.}~\cite{ref1,ref2} have
developed a model based on the addition of the different effects
assuming that all effects are small over one round-trip of the
cavity. Analytical studies of Akhmediev {\it et al.}~\cite{ref16,ref17} are based
on a normalized complex cubic Ginzburg-Landau (CGL) equation and
give the stability conditions of the mode-locked solutions.
On
the other hand, many numerical simulations have been done to
complete analytic approaches \cite{ref18}-\cite{ref20}.
We have recently investigated experimentally and theoretically the mode-locking properties of an Yb-doped double clad fiber
laser passively mode-locked through nonlinear polarization rotation \cite{ref12,ref21}. The optical configuration was a
unidirectional ring cavity containing an optical isolator placed between two halfwave plates. Only two phase plates were
considered for simplicity. The theoretical model reduces to a complex cubic Ginzburg-Landau equation whose coefficients
explicitly depend on the orientation of the phase plates. The model allowed the description of both the self-starting
mode-locking operation and the operating regimes as a function of the orientation of the halfwave plates. The model was
then adapted to the anomalous dispersion case \cite{ref22} and to the stretched-pulse operation \cite{ref23}. Although
our simplified model is in good agreement with the experimental results, a typical experiment includes two polarization
controllers instead of two halfwave plates. Indeed, mode-locking is more easily obtained
 in the former case because there is
more degrees of freedom. The aim of this paper is to provide a general model taking into account a polarizer and two sets
of a halfwave plate and a quarterwave plate. The paper is organized as follows. In section 2 we derive a propagation
equation for a unidirectional ring cavity containing a polarizer placed between two sets of a halfwave and a quarterwave
plates. The resulting equation, valid for a large number of round trips, is of the CGL type and explicitly takes into
account the orientation of the phase plates and the polarizer. Constant and mode-locked solutions are considered in
section 3. The last section is devoted to  a discussion of
 the influence of the phase plates and of the polarizer on the stability
of both the mode-locking and the continuous wave regimes of the laser.
\begin{figure}[hbt!]
\begin{center}
\includegraphics[width=9cm]{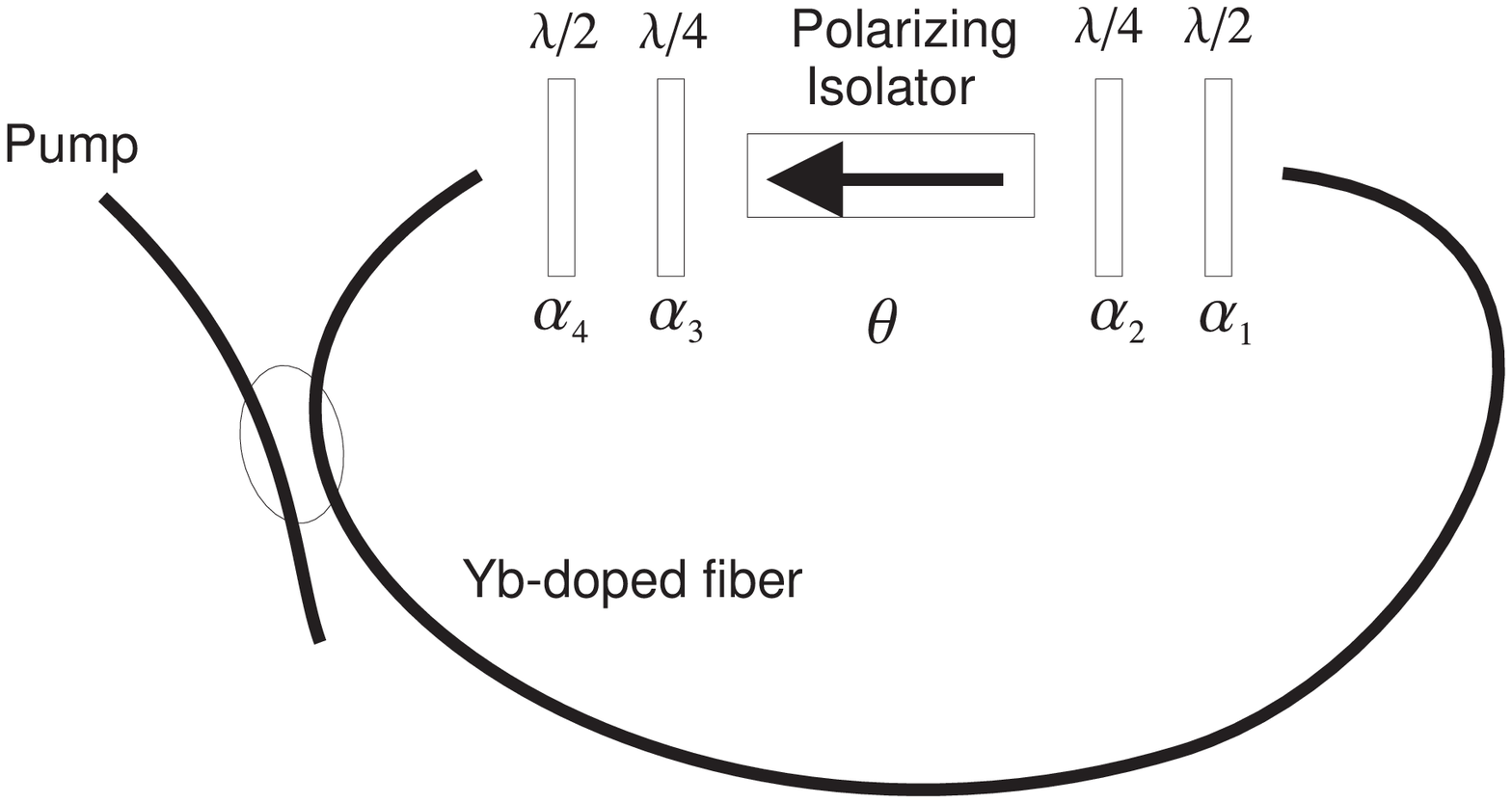}
\caption{\footnotesize Schematic representation of the fiber laser passively mode-locked through nonlinear polarization
rotation.}
\label{montage}
\end{center}
\end{figure}

\section{The master equation}

In this section we derive a master equation for the laser shown in figure~\ref{montage}. The ytterbium-doped fiber has
gain, birefringence, group velocity dispersion (GVD) and optical Kerr nonlinearity. The cavity contains a
polarizing isolator placed between two polarization controllers.

\subsection{Propagation along the ytterbium-doped fiber}

In the framework of the eigenaxis of the fiber moving at the group velocity, the
propagation equations for the two polarization components of the
amplitude of the electric field are \cite{ref12,ref24,ref25}
\bq i\frac{\partial u}{\partial z}-K u-\frac{\beta_2}2\frac{\partial^2 u}{\partial t^2}
+\gamma\left(u\left|u\right|^2+Au\left|v\right|^2+Bv^2u^\ast\right)=
ig\left(1+\frac1{\omega_g^2}\frac{\partial^2 }{\partial t^2}\right)u,\label{1}\eq
\bq i\frac{\partial v}{\partial z}+K v-\frac{\beta_2}2\frac{\partial^2 v}{\partial t^2}
+\gamma\left(v\left|v\right|^2+Av\left|u\right|^2+Bu^2v^\ast\right)=
ig\left(1+\frac1{\omega_g^2}\frac{\partial^2 }{\partial t^2}\right)v,\label{2}\eq
where
$g$ in $\rm m^{-1}$ is the linear gain,
$\omega_g=10^{13}\;\rm s^{-1}$ the spectral gain bandwidth, $A=2/3$, and $B=1/3$. $K$ is the birefringent parameter and
$\gamma$ the nonlinear coefficient.\\
Following our analysis of reference \cite{ref12}, we assume that the effects of the GVD $\beta_2$, the nonlinear effect
$\gamma$, and the gain filtering $\rho={g}/{\omega_g^2}$ are small over one round-trip of the cavity.
A perturbative approach
can be used. We introduce a small parameter
$\e$ and replace the quantities $\beta_2$, $\gamma$ and $\rho$ by $\e\beta_2$, $\e\gamma$ and $\e\rho$. Let ($u(0),v(0)$)
the electric field components at the entrance of the ytterbium-doped fiber, and ($u(L),v(L)$) the components at the exit
of the fiber of length $L$. A first order perturbative calculation leads to~\cite{ref12,ref22}
\bq \ba{rr} \displaystyle u(L)=& \displaystyle
u(0)e^{(g-iK)L}+\e\biggl[
L\left(\rho-\frac{i\beta_2}2\right)\frac{\partial^2 u(0)}{\partial t^2}
\hspace{4cm}\\
&\displaystyle+i\gamma\left(u(0)\left|u(0)\right|^2
+A u(0)\left| v(0)\right|^2\right)\frac{e^{2gL}-1}{2g}
\hspace{3cm}\\
&\displaystyle
+i\gamma B v(0)^2 u(0)^\ast\frac{e^{(2g+4iK)L}-1}{2g+4iK}\biggr]e^{(g-iK)L}
+O\left(\e^2\right),\ea
\label{3}\eq
\bq \ba{rr}
\displaystyle v(L)=&
\displaystyle v(0)e^{(g+iK)L}+\e\biggl[ L\left(\rho-\frac{i\beta_2}2\right)
\frac{\partial^2 v(0)}{\partial t^2}
\hspace{4cm}\\
&\displaystyle+i\gamma\left(v(0)\left|v(0)\right|^2
+A v(0)\left| u(0)\right|^2\right)\frac{e^{2gL}-1}{2g}
\hspace{3cm}\\
&\displaystyle
+i\gamma B u(0)^2 v(0)^\ast\frac{e^{(2g-4iK)L}-1}{2g-4iK}\biggr]e^{(g+iK)L}
+O\left(\e^2\right).\ea
\label{4}\eq

\subsection{Modelling the phase plates and the polarizer}

The Jones matrix formalism is well adapted to the treatment of a combination of phase plates and polarizer.
It will be used in this section. Without loss of generality, we assume that the eigenaxis at both ends of the
fiber are aligned and parallel to the $x$ and $y$-axes of the laboratory frame. Let $\alpha_1$ (resp. $\alpha_4$)
the angle between the eigenaxis of the halfwave plate and the $x$-axis before (resp. after) the polarizer.
Let $\alpha_2$ (resp. $\alpha_3$) the angle between the eigenaxis of the quarterwave plate and the $x$-axis
before (resp. after) the polarizer. Let $\theta$ the angle between the passing axis of the polarizer and the $x$-axis.

In the framework of their eigenaxis, the Jones matrices of the quarterwave and halfwave plates are respectively
\bq M_{\frac{\lambda}{4}}=\frac{\sqrt 2}2 \left(\ba{c}1-i\\0\ea \ba{c}0\\1+i\ea\right),\label{5}\eq
\bq M_{\frac{\lambda}{2}}=\left(\ba{c}-i\\0\ea \ba{c}0\\i\ea\right).\label{6}\eq
Let $M_3$ (resp. $M_4$) be the Jones matrix of the quarterwave plate (resp. halfwave plate) after the isolator
in the $(Ox,Oy)$
frame:
\bq M_3=R(\alpha_3)M_{\frac{\lambda}{4}}R(-\alpha_3),\label{7}\eq
\bq M_4=R(\alpha_4)M_{\frac{\lambda}{2}}R(-\alpha_4),\label{8}\eq
where
\bq R(\alpha)=\left(\ba{c}\cos\alpha\\\sin\alpha\ea \ba{c}-\sin\alpha\\\cos\alpha\ea\right)\label{9}\eq
is the rotation matrix of angle $\alpha$.

Light exiting the polarizer passes through a set of a quarterwave and a halfwave plates. Therefore the electric field at
the entrance of the fiber after the $n^{\rm th}$ round trip is
\bq\left(\ba{c}u_n{(0)}\\v_n{(0)}\ea\right)=M_4M_3\left(\ba{c}u'_n\\v'_n\ea\right),\label{10}\eq
where $ u'_n$ and $ v'_n$ are the electric field components just after the polarizer.

Let $M$ be the Jones matrix of the polarizer and $M_1$ (resp. $M_2$)
the Jones matrix of the halfwave plate (resp. quarterwave) before the polarizer. In the $(Ox,Oy)$
frame, the matrices write as
\bq M=R(\theta)\left(\ba{c}\beta\\0\ea \ba{c}0\\0\ea\right)R(-\theta),\label{11}\eq
where $\beta=95\%$ is the transmission coefficient of the polarizer, and
\bq M_1=R(\alpha_1)M_{\frac{\lambda}{2}}R(-\alpha_1),\quad M_2=R(\alpha_2)M_{\frac{\lambda}{4}}R(-\alpha_2).\label{13}\eq

The field after the polarizer can be written as
\bq\left(\ba{c}u'_{n+1}\\v'_{n+1}\ea\right) =\left(\ba{c}\cos\theta\\\sin\theta\ea\right)f_{n+1}
=M M_2 M_1\left(\ba{c}u_n{(L)}\\v_n{(L)}\ea\right),\label{14}\eq
where $f_{n+1}$ is the electric field amplitude after the polarizer at the $(n+1)^{\rm th}$ round trip.

We now replace the matrices $M$, $M_1$, and $M_2$ by expressions (\ref{11}),  and (\ref{13})
respectively. We further take for $\left(u_n(L),v_n(L)\right)$ the expressions given in (\ref{3},\ref{4}),
 and $\left(u_n(0),v_n(0)\right)$ is replaced by equation (\ref{10}).
  Finally, we take into account equations (\ref{7}) and (\ref{8}), and
   get a relation between $f_{n+1}$ and $f_{n}$:
\bq f_{n+1}=\beta e^{gL}\left\{Q f_n+\e\left[\left(\rho-\frac{i \beta_2}2\right) L Q\;
\frac{\partial^2 f_n}{\partial t^2} +i P
f_n\left|f_n\right|^2\right]\right\}+O\left(\e^2\right),\label{15}\eq
where the coefficient $P$ and $Q$ are given in the appendix.
The important fact in our analysis is that coefficients $P$ and $Q$ explicitly
 depend on the angles $\alpha_1$, $\alpha_2$, $\alpha_3$, $\alpha_4$, and $\theta$.
 As we will see in the next section, the model will allow to investigate the operating regime of the laser
  as a function of the orientation of the phase plates and of the polarizer.

A stationary state is reached when $|f_{n+1}|$=$|f_{n}|$.
This occurs when the gain attains its threshold value $ g=g_0+\e g_1+O\left(\e^2\right)$.
$g_1$ is referred to  as the excess of linear gain below.
 The dominant part of $f_{n+1}$ is obtained at order $\e^0$:
\bq f_{n+1}=\beta e^{g_0L}Q f_n+O\left(\e\right).\label{16}\eq
As a consequence of the stationarity, the modulus of $\beta e^{g_0L}Q$ is unity. We thus obtain the expression of $g_0$,
as
\bq\ba{rl} g_0=&\displaystyle\frac{-1}{2L}\ln\left(\beta^2\left|Q\right|^2\right)\vspace{1.5mm}\\
=&\displaystyle \frac{-1}{2L}\ln\left(\beta^2\left[\vert\phi_1\vert^2+e^{2iKL}\phi_1^\ast\phi_2+
e^{-2iKL}\phi_1\phi_2^\ast+\vert\phi_2\vert^2\right]\right).\ea\label{17}\eq

By performing a Taylor expansion of $e^{\e g_1 L}$,
 and replacing $\beta e^{g_0 L} Q$ by $e^{i\psi}$, equation (\ref{15}) becomes
\bq f_{n+1}=e^{i \psi}\left(1+\e g_1L\right)f_n+\e\left(\rho-\frac{i\beta_2}2 \right)Le^{i \psi}
\frac{\partial^2 f_n}{\partial t^2} +i
\e \frac{e^{i\psi}}Q P f_n\left|f_n\right|^2+O\left(\e^2\right).\label{19}\eq
It is more convenient to describe the evolution of the field amplitude $f_n$ by a continuous equation.
The discrete sequence   $f_n$ is interpolated by a continuous function and,
for a large number of round trips $n\propto 1/\e$, a fast rotating phase factor is set apart~\cite{ref12,ref22},
 which yields the equation
\bq i\frac{\partial F}{\partial \zeta}=i g_1 F+\left(\frac{\beta_2}2+i\rho\right)
\frac{\partial^2 F}{\partial t^2}+
\left({\cal D}_r+i{\cal D}_i\right)F\left|F\right|^2,\label{23}\eq
where
\bq F\left(\zeta=\e n L\right)=f_n e^{-in\psi}+O\left(\e\right),\label{22}\eq
and ${\cal D}_r$ and ${\cal D}_i$ are the real and imaginary parts of the
quantity ${\cal D}$ given by
\bq{\cal D}=\frac{-P}{QL}. \label{21}\eq
 They correspond respectively to the effective self-phase modulation and to the effective nonlinear gain or absorption.
 ${\cal D}_r$ is always negative while the sign of ${\cal D}_i$ depends on $\alpha_1$, $\alpha_2$, $\alpha_3$, $\alpha_4$,
 and $\theta$. Equation (\ref{23}) is of cubic complex Ginzburg-Landau type (CGL).

\section{Solution of the CGL equation}

This section is devoted to the study of two particular solutions of equation~(\ref{23}).
 We first consider the constant solution corresponding to a continuous wave (CW)
  operating regime of the laser. Localized solutions are then considered and are related to the
   mode-locking regime of the laser. In both cases, the stability criterium of the solution is determined.

\subsection{Constant amplitude solution}

A constant amplitude solution of CGL is
\bq F={\cal A} e^{i\left(k\zeta-\Omega t\right)},\label{38}\eq
where
\bq \Omega^2=\frac{1}{\rho}\left({\cal D}_i\vert{\cal A}\vert^2+g_1\right),
\qquad
 k=\frac{\beta_2}{2\rho}\left({\cal D}_i\vert {\cal A}\vert^2+g_1\right)-{\cal D}_r\vert{\cal A}\vert^2.\label{40}\eq
Solution (\ref{38}) is time independent if $\Omega=0$. Under this condition, the expressions of ${\cal A}$ and $k$ are
\bq {\cal A}=\sqrt{\frac{-g_1}{{\cal D}_i}}, \qquad k=\frac{{\cal D}_r }{{\cal D}_i}g_1.\label{42}\eq
This solution exists only if ${\cal D}_ig_1$ is negative.
 On the other hand, it has been demonstrated that the modulational instability occurs when the excess
 of linear gain $g_1$ is negative and the effective nonlinear gain ${\cal D}_i$ is positive \cite{ref12}.
  Therefore the constant amplitude solution is stable when the excess of linear gain is positive and the
   effective nonlinear gain ${\cal D}_i$ is negative.

\subsection{Localized solution}

Equation (\ref{23}) admits the following localized solution:
\bq F=a(t)^{1+id}e^{-i\omega\zeta},\label{43}\eq
where
\bq d=\frac{-3\left[\beta_2{\cal D}_r+2\rho{\cal D}_i\right]+\sqrt{9\left[2\rho{\cal D}_i+\beta_2{\cal D}_r\right]^2+8
\left[\beta_2{\cal D}_i-2\rho{\cal D}_r\right]^2}}{2\left[\beta_2{\cal D}_i-2\rho{\cal D}_r\right]},\label{44}\eq
\bq\omega=\frac{-g_1 \left[4\rho d+\beta_2 d^2-\beta_2\right]}{2\left[\rho d^2-\rho-\beta_2 d\right]}.\label{45}\eq
The  parameter $d$ represents the chirp. The amplitude $a(t)$ writes as
\bq a(t)=MN\,{\rm sech}\,(Mt),\label{46}\eq
where
\bq M=\sqrt{\frac{g_1}{\rho d^2-\rho-\beta_2 d}}\;,\label{47}\eq
\bq N=\sqrt{\frac{3d\left[4\rho^2+\beta_2^2\right]}{2\left[\beta_2{\cal D}_i-2\rho{\cal D}_r\right]}}\;.\label{48}\eq
The pulses exist if both $M$ and $N$ are real. Stability of the localized solution results from an equilibrium between
the excess of linear gain, the quantity $\beta_2{\cal D}_r$, and the effective nonlinear gain. Indeed, in the
defocusing case where $\beta_2{\cal D}_r<0$, the pulse is potentially stable if the excess of linear gain $g_1$
is negative and the effective nonlinear gain ${\cal D}_i$ is positive. This criterium can be written in the
mathematical form \cite{ref12}
\bq \left(\rho d^2-\rho-\beta_2 d\right)<0.\label{49}\eq
When the effective nonlinear gain is negative, the stability of the pulses is not known at this time.
Note that higher order terms or gain saturation can definitely stabilize the short pulse solution of equation (\ref{23}).

\section{Influence of the orientations of the phase plates and of the polarizer}

In the previous section we have derived a master equation for a
laser passively mode-locked by nonlinear polarization rotation.
The coefficients of the equation depend on the orientation angles
of the phase plates $\alpha_1$, $\alpha_2$, $\alpha_3$,
$\alpha_4$, and of the polarizer $\theta$. As a consequence, the
stability of both the continuous and the mode-locked solutions
also depends on these angles. Because of the large number of
degrees of freedom, we cannot perform a systematic study of the
stability of the solutions as a function of the five angles. In
the following we have generally fixed three angles and varied the
two remaining ones. In these conditions it is convenient to
summarize the results in a two dimensional stability diagram
which gives for any couple of varying angles the regions of
stability of both the continuous and the mode-locked solutions. We
have first considered $(\theta,\alpha_2,\alpha_3)=(\theta,0^\circ,0^\circ)$ where
$\theta$ takes the following values: $0^\circ$, $30^\circ$, $45^\circ$,
$60^\circ$, $90^\circ$, $120^\circ$, $135^\circ$, $150^\circ$, and $180^\circ$. We have
plotted the stability diagram in the plane $(\alpha_1,\alpha_4)$
for each value of $\theta$. The same studies have been done for
$(\theta,\alpha_2,\alpha_3)=(0^\circ, \alpha_2,0^\circ)$,
$(0^\circ,0^\circ,\alpha_3)$, $(30^\circ,30^\circ,30^\circ)$,
$(45^\circ,120^\circ,150^\circ)$, and $(60^\circ,30^\circ,135^\circ)$.
In the two first cases, $\alpha_2$ and
$\alpha_3$ take the same values as  attributed to $\theta$.
For the numerical computations, we have used the same parameters
as in ref.~\cite{ref12}: $K=1.5\;\rm m^{-1}$, $\beta_2=0.026\;
\rm ps^{2}m^{-1}$, $L=9\;\rm m$ and $\gamma=3\cdot10^{-3}\;\rm W^{-1} m^{-1}$.\\

\begin{figure}[hbt!]
\begin{center}
\includegraphics[width=9cm]{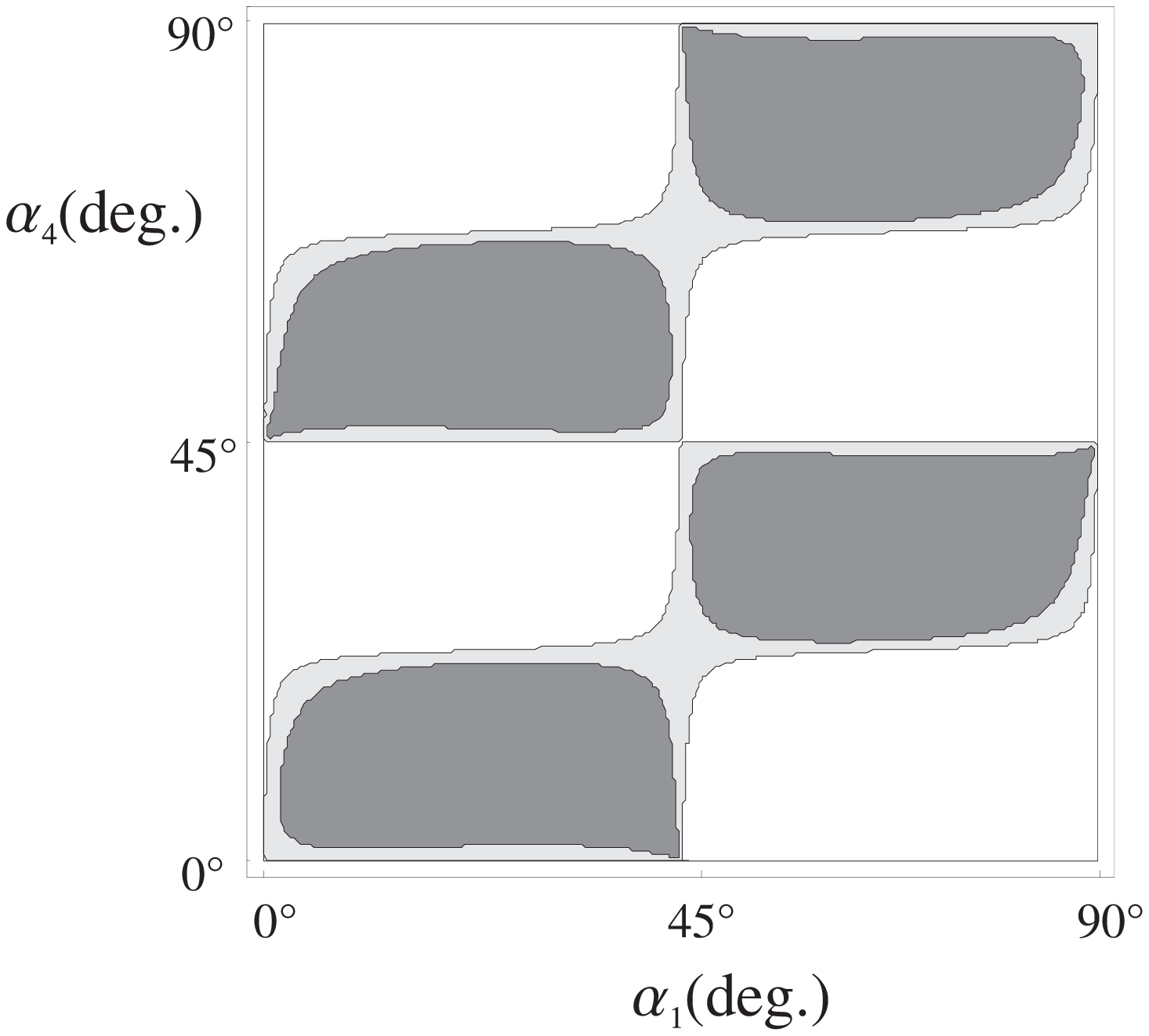}
\caption{\footnotesize Stability diagram of the CW and the mode-locked solutions in the plane
($\alpha_1,\alpha_4$) for $(\theta,\alpha_2,\alpha_3)=(0^\circ,0^\circ,0^\circ)$.
The white region corresponds to stable CW
operation and unstable mode-locking, the light gray corresponds to unstable CW and unstable mode-locking and the
dark gray region corresponds to stable mode-locking operation and unstable CW. }
\label{carte1}
\end{center}
\end{figure}

\begin{figure}[hbt!]
\begin{center}
\includegraphics[width=9cm]{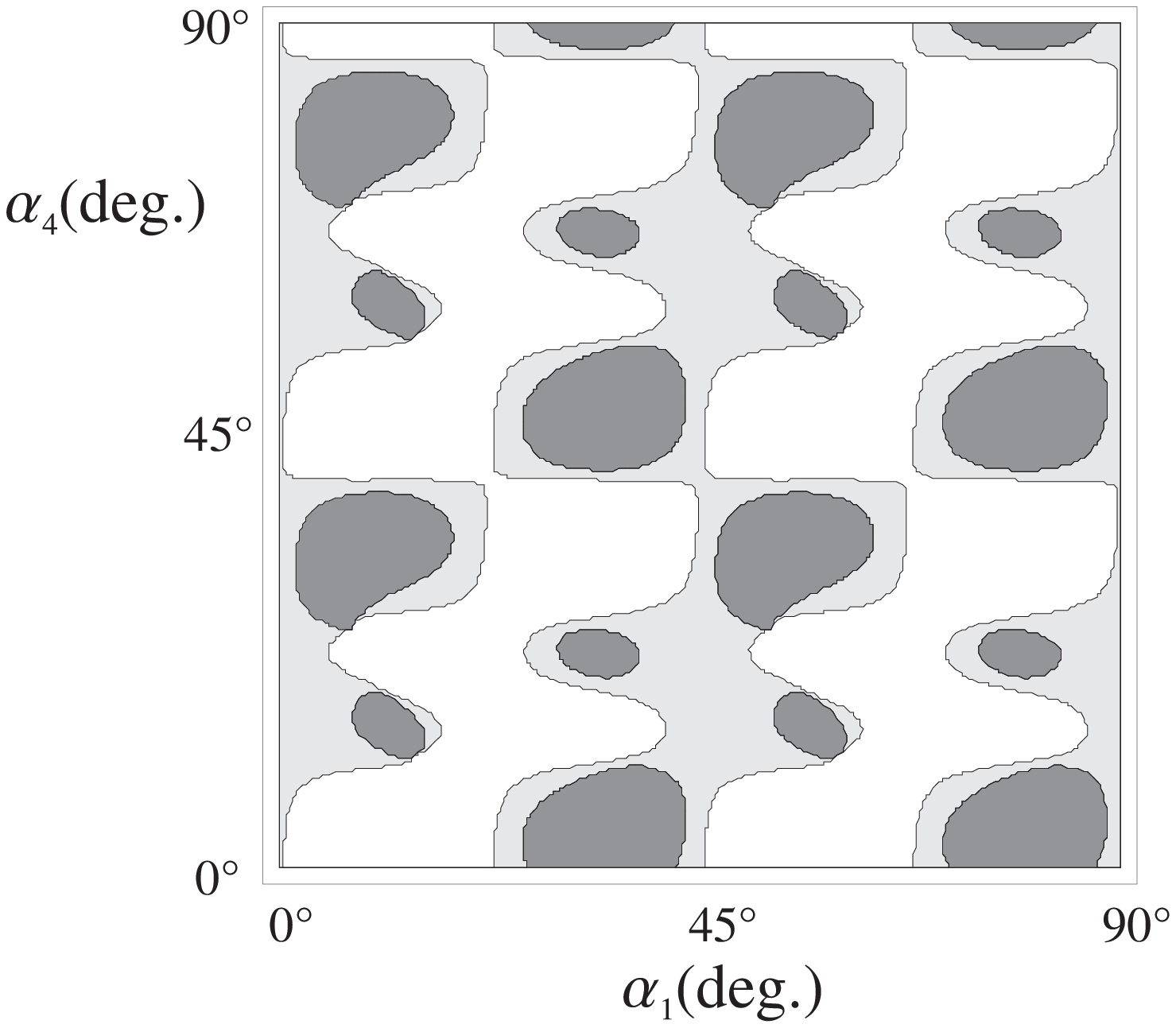}
\caption{\footnotesize Stability diagram of the CW and the mode-locked solutions in the plane
($\alpha_1,\alpha_4$) for $(\theta,\alpha_2,\alpha_3)=(0^\circ,0^\circ,30^\circ)$.
The white region corresponds to stable CW
operation and unstable mode-locking, the light gray corresponds to unstable CW and mode-locking and the
dark gray region corresponds to stable mode-locking operation and unstable CW. }
\label{carte18}
\end{center}
\end{figure}

\begin{figure}[hbt!]
\begin{center}
\includegraphics[width=9cm]{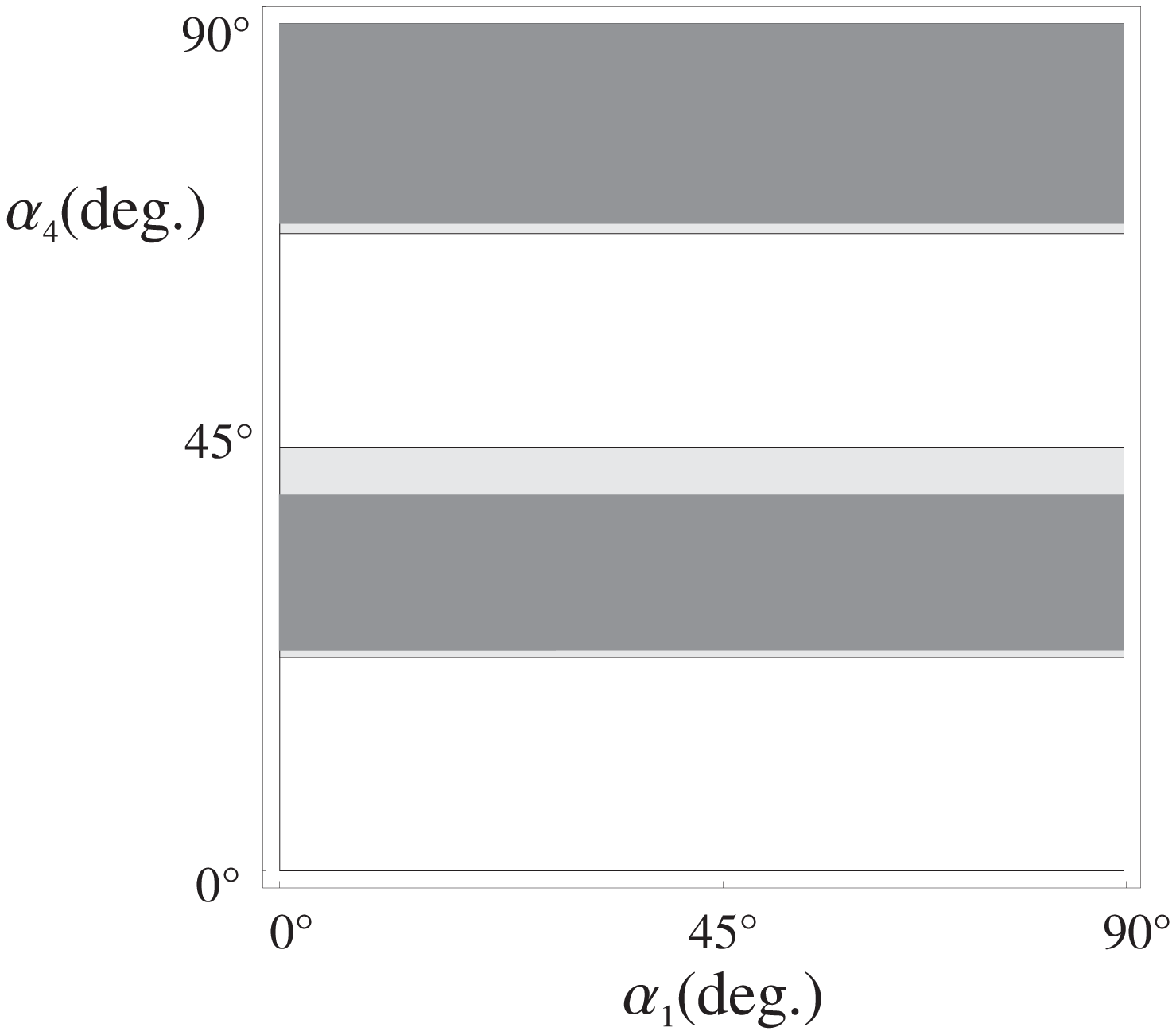}
\caption{\footnotesize Stability diagram of the CW and the mode-locked solutions in the plane
($\alpha_1,\alpha_4$) for $(\theta,\alpha_2,\alpha_3)=(0^\circ,45^\circ,0^\circ)$.
The colors have the same meaning
as in figure \ref{carte1}.}
\label{carte2}
\end{center}
\end{figure}

\begin{figure}[hbt!]
\begin{center}
\includegraphics[width=9cm]{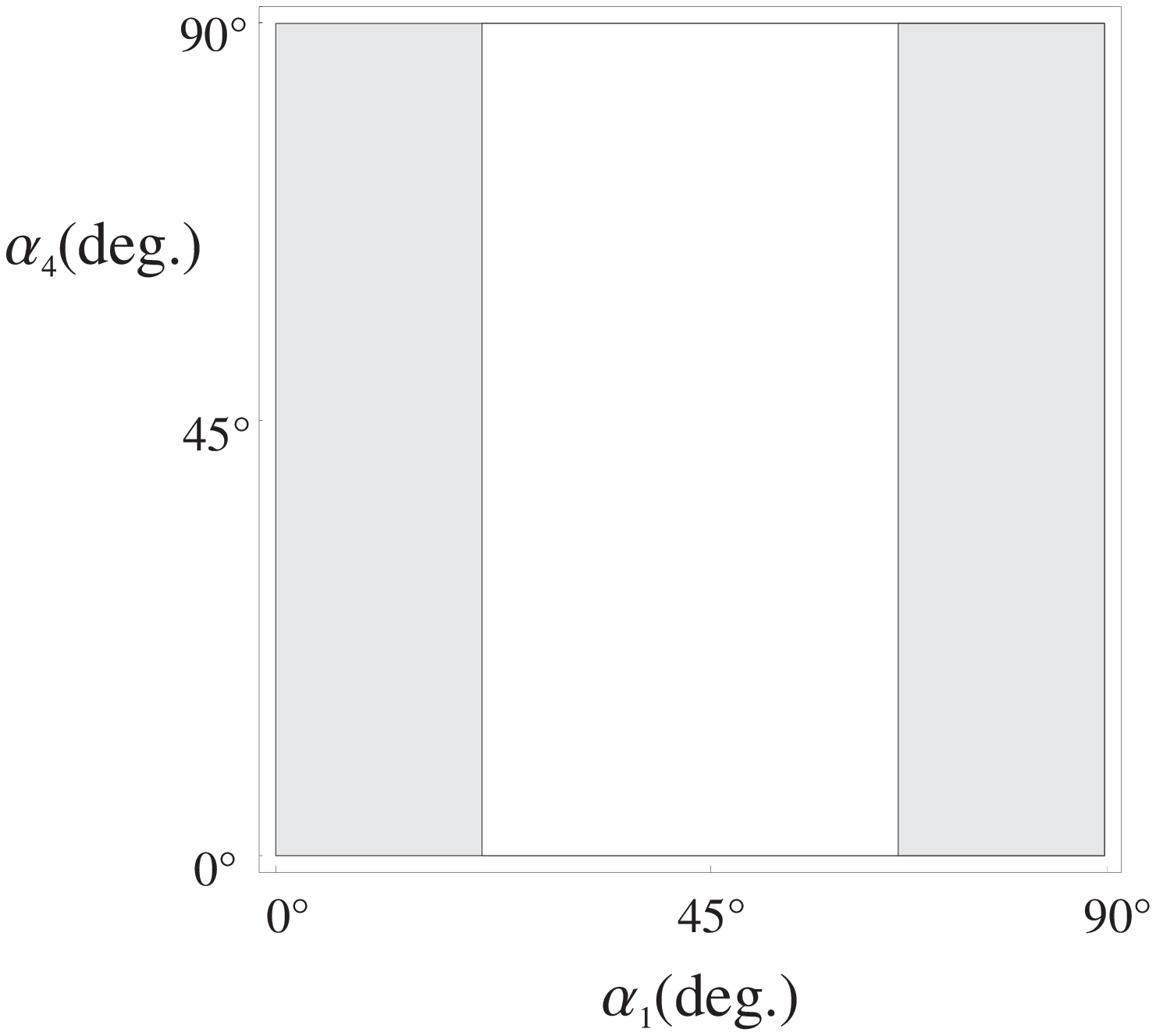}
\caption{\footnotesize Stability diagram of the CW and the mode-locked solutions in the plane ($\alpha_1,\alpha_4$)
for $(\theta,\alpha_2,\alpha_3)=(0^\circ,0^\circ,45^\circ)$.
 The colors have the same meaning as in figure~\ref{carte1}.}
\label{carte3}
\end{center}
\end{figure}

A great dependance of the stability domains versus $\alpha_1$,
$\alpha_2$, $\alpha_3$, $\alpha_4$, and $\theta$ have been
observed. This can be physically expected because a change in the
orientation of one element leads to a relative variation of the
losses undergo by the wings and the center of the pulse. It is
then possible either to favor the center of the pulse which
travels the polarizer with a minimum losses, leading to efficient
mode-locking regime, or to favor the opposite case resulting in
the instability of the mode-locking regime. These results are
illustrated in figures \ref{carte1}, \ref{carte18}, \ref{carte2}, and
\ref{carte3}.
 They give the stability domains of the CW and
mode-locking regimes depending on the orientation angles $(\alpha_1,\alpha_4)$
of the halfwave plates,
for the following orientations of the polarizer and quarterwave plates:
$(\theta,\alpha_2,\alpha_3)=(0^\circ,0^\circ,0^\circ)$, $(0^\circ,0^\circ,30^\circ)$,
$(0^\circ,45^\circ,0^\circ)$, and
$(0^\circ,0^\circ,45^\circ)$, respectively. The representations
have been limited to $0^\circ\leqslant\alpha_1,\alpha_4\leqslant90^\circ$
because of the periodicity. Figure \ref{carte1} is the same that the one in
reference \cite{ref12} where only two halfwave plates were
considered. This is correct because the polarizer is aligned with
the eigenaxis of the two quarterwave plates. Thus this result
validates the general model including four phase plates.
A large part of the computed cartographies are relatively close to figure \ref{carte1},
but another typical shape is shown on figure \ref{carte18}.
 Figures
\ref{carte2} and \ref{carte3} show that the operating regime can
be independent of the orientation of one of the halfwave plates.
We can note on figure \ref{carte3} that the orientation of the last half wave
plate ($\alpha_4$) does not modify the stability of the solutions in
this case. This is not surprising since for $\theta=0^\circ$ and $\alpha_3=45^\circ$,
the polarization that enters this last half wave plate is circular.
Whatever the orientation of this plate, the polarization
entering the fiber is thus circular, and the global behavior does not
depend on $\alpha_4$. This further allows to give a physical
interpretation to the absence of any mode-locking domain in this
case. We can see from relations (3) and (4) that in absence of
birefringence ($K=0$), if a circular polarization enters the fiber,
a circular polarization exits the fiber. Actually, nonlinear
polarization rotation does not occur. We can thus assume that this is the
reason why no mode-locking regime is predicted here.

\begin{figure}[hbt!]
\begin{center}
\includegraphics[width=9cm]{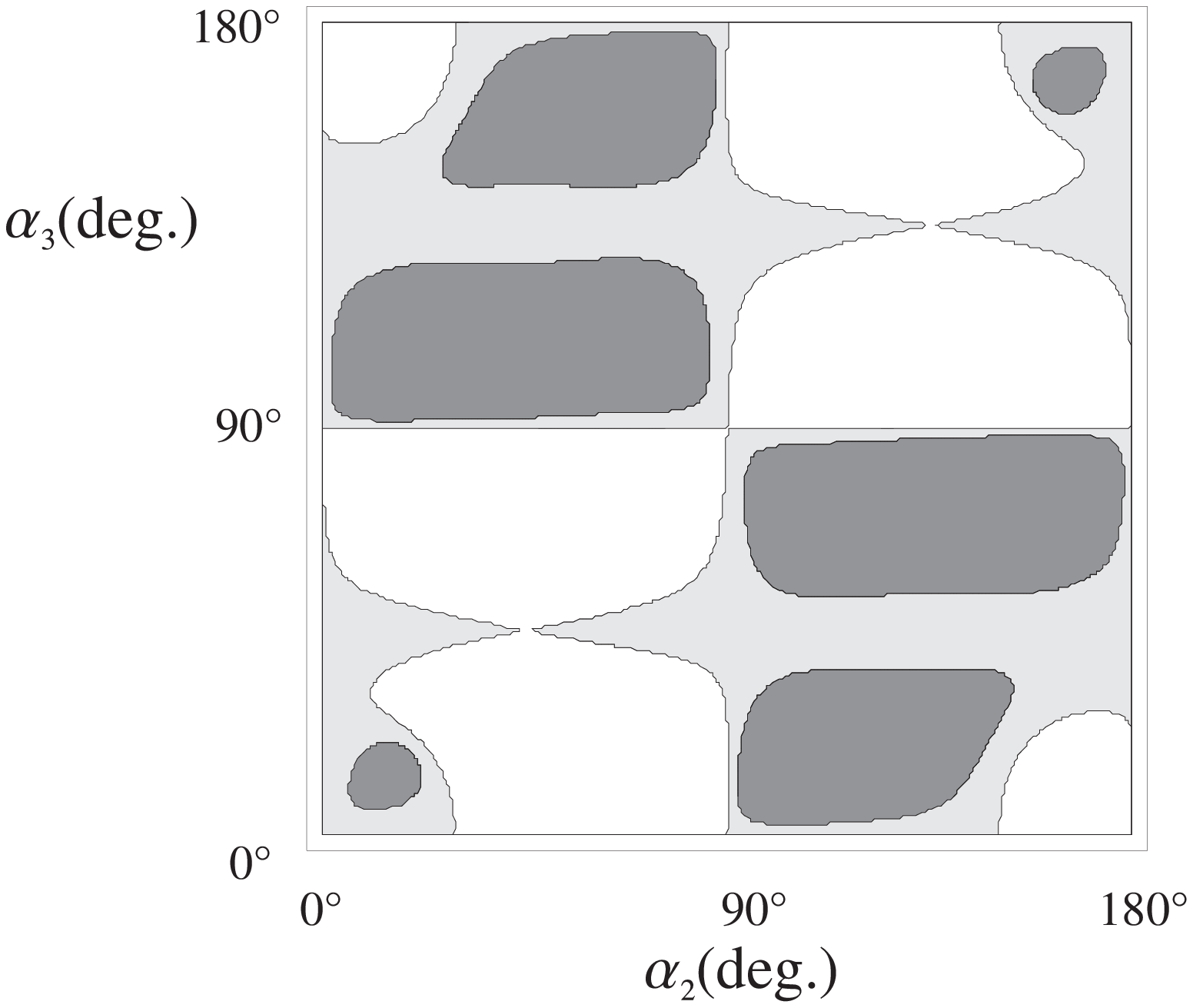}
\caption{\footnotesize Stability diagram of the CW and the mode-locked solutions in the plane $(\alpha_2,\alpha_3)$
for $(\theta,\alpha_1,\alpha_4)=(0^\circ,0^\circ,0^\circ)$. The colors have the same meaning as in figure~\ref{carte1}.}
\label{carte19}
\end{center}
\end{figure}

\begin{figure}[hbt!]
\begin{center}
\includegraphics[width=9cm]{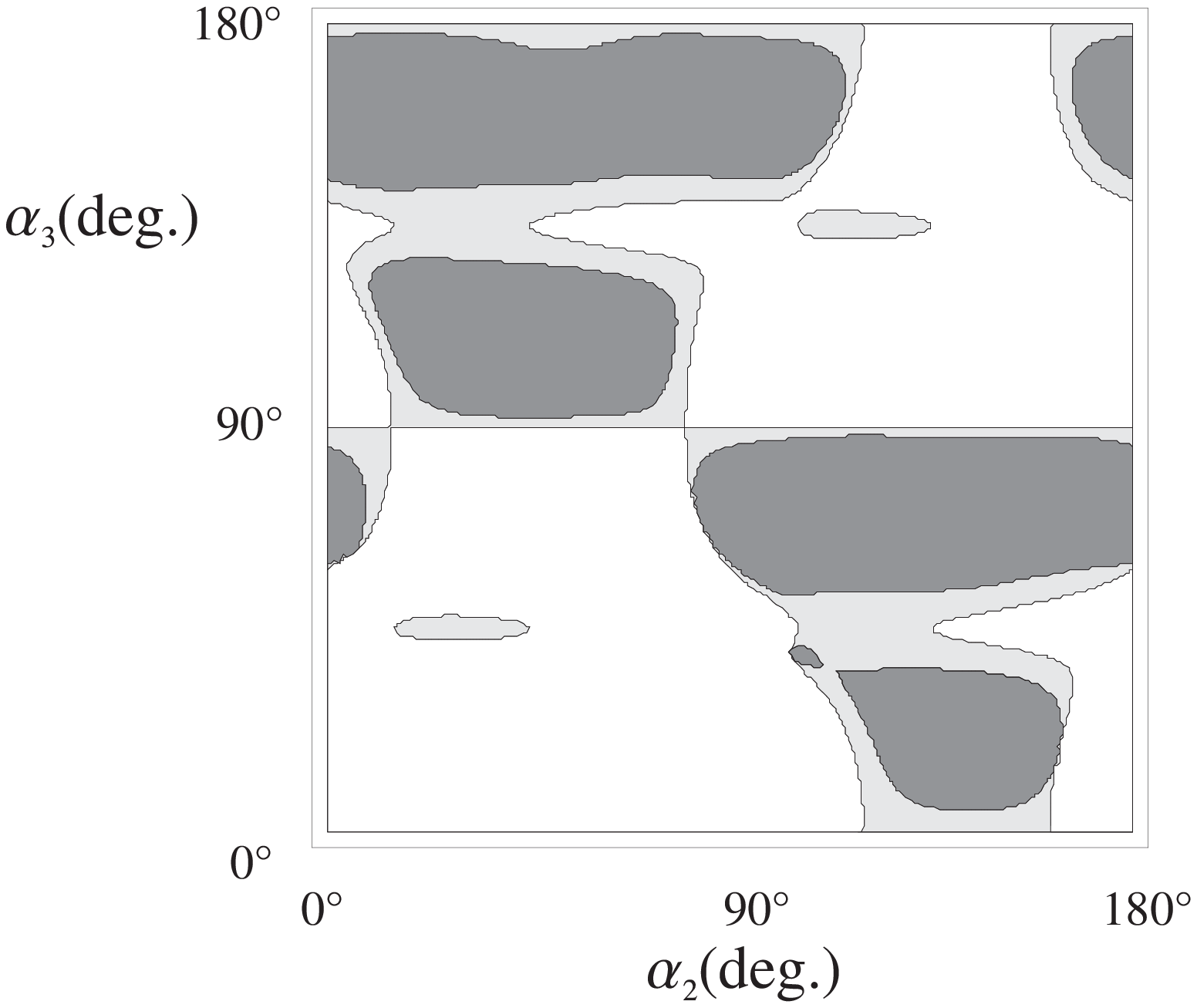}
\caption{\footnotesize Stability diagram of the CW and the mode-locked solutions in the plane $(\alpha_2,\alpha_3)$
for $(\theta,\alpha_1,\alpha_4)=(0^\circ,30^\circ,0^\circ)$. The colors have the same meaning as in figure~\ref{carte1}.}
\label{carte20}
\end{center}
\end{figure}

\begin{figure}[hbt!]
\begin{center}
\includegraphics[width=9cm]{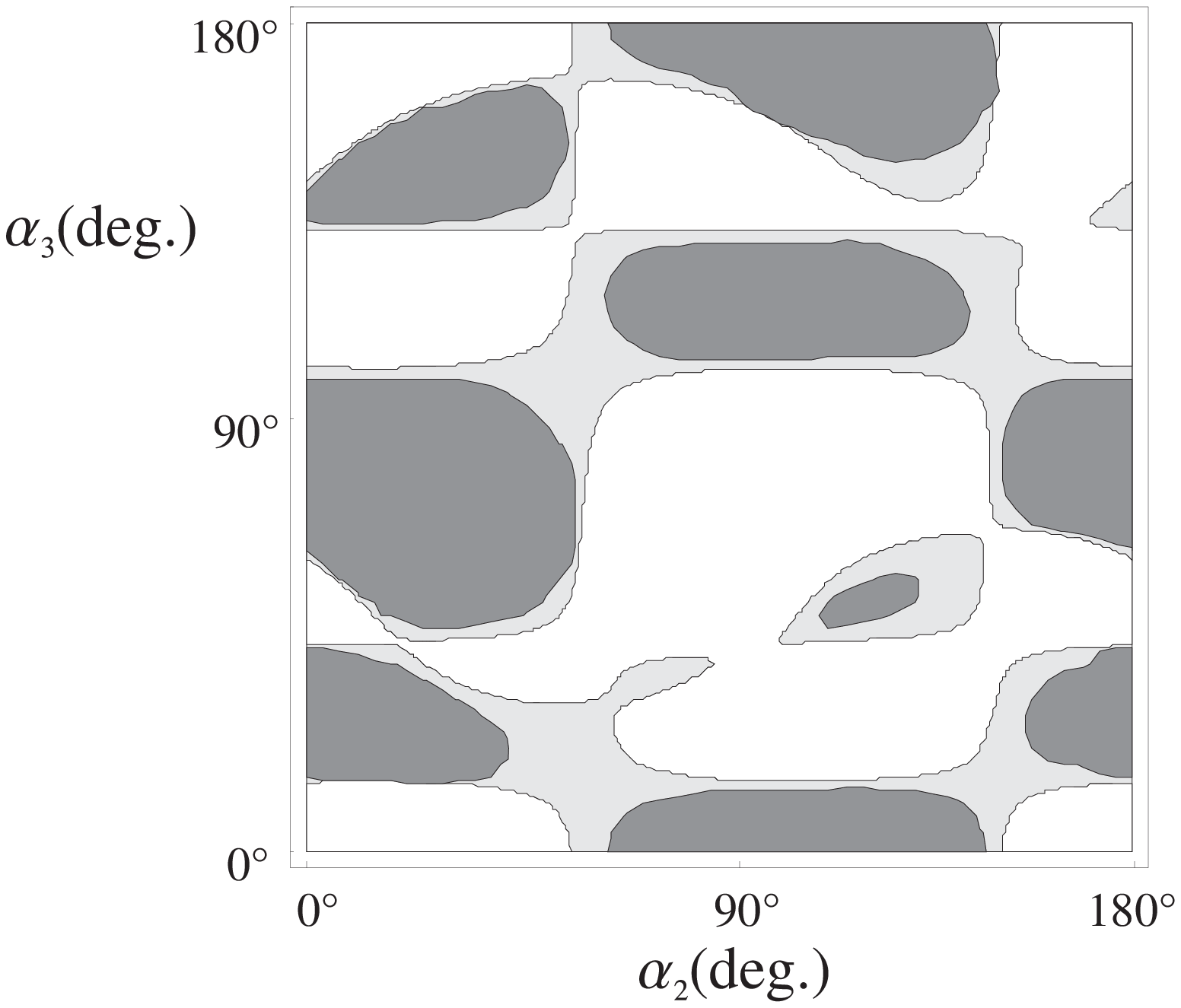}
\caption{\footnotesize Stability diagram of the CW and the mode-locked solutions in the plane $(\alpha_2,\alpha_3)$
for $(\theta,\alpha_1,\alpha_4)=(60^\circ,30^\circ,135^\circ)$. The colors have the same meaning as in figure~\ref{carte1}.}
\label{carte4}
\end{center}
\end{figure}

We have then explored the dependency of the operating regimes of the laser
with respect to the orientation angles $(\alpha_2,\alpha_3)$
of the quarterwave plates. The periodicity versus $\alpha_2$ and
$\alpha_3$ is $180^\circ$. Figures \ref{carte19}, \ref{carte20} and \ref{carte4} give
typical   examples of cartographies.
They are obtained  for the orientations
 $(\theta,\alpha_1,\alpha_4)=(0^\circ,0^\circ,0^\circ)$, $(0^\circ,30^\circ,0^\circ)$,
 and $(60^\circ,30^\circ,135^\circ)$
of the polarizer and halfwave plates. We can note
the large regions of instability and also the increased number of
mode-locking regions compared to the reference results of figure
\ref{carte1}, especially on figure \ref{carte4}. It is interesting to
point out the existence of four horizontal axes that separate
abruptly the different domains and where no mode-locking is
observed. They locate at values of $\alpha_3$ about integer multiples
of $45^\circ$, on figures \ref{carte19}-\ref{carte20}, and  around $15^\circ$, $45^\circ$,
$105^\circ$ and $135^\circ $ on figure \ref{carte4}. In the latter case, $\theta=60^\circ$,
while it is zero in the former. We
can thus deduce that for $\alpha_3=60^\circ \pm 45^\circ $,
polarization exiting the plate $\rm n^o 3$  is circular, which is
not modified by the last plate $\rm n^o 4$ ($\lambda/2$). As previously, we can
assume that nonlinear polarization rotation does not occur such
that mode-locking is not observed. These cases correspond indeed
to the horizontal axes where $\alpha_3$ is around
$45^\circ$ or $135^\circ$ on figures \ref{carte19}-\ref{carte20},
$15^\circ$ or
$105^\circ$ on figure \ref{carte4}. In addition, these axes appear as boundaries: when
$\alpha_3$ passes through these axes, the ratio between the
$x$-polarized and the $y$-polarized components entering the fiber
passes unity, ``inverting" the effect of nonlinear polarization
rotation and thus on mode-locking or CW operation. We have checked
with other values of $\theta$ the existence of similar horizontal
axes at $\alpha_3=\theta \pm 45^\circ$ , that separate abruptly
mode-locking and CW domains and where mode-locking does not occur in
general. Other axes, around $\alpha_3=0^\circ$ and $90^\circ$
on figures \ref{carte19}-\ref{carte20} or $45^\circ$  and $135^\circ$ on figure \ref{carte4},
can be interpreted with similar arguments. The eigenaxes of this
plate are then parallel to those of wave plate $\rm n^o 4$
($\alpha_4=0^\circ$ in the former case, $135^\circ$ in the latter).
 Then the polarization entering the
fiber is in general elliptical, but with its high-axis oriented at
$45^\circ $ from the $x$-axis and $y$-axis of the fiber. The maximum of
$x$ and $y$ amplitudes in the fiber are thus identical and we can
assume that nonlinear polarization rotation is not efficient. To
confirm this assumption, we have plotted another cartography in
the ($\alpha_2,\alpha_3$) plane with the same parameters:
$\theta=30^\circ $ and $\alpha_1=30^\circ $, but with
$\alpha_4=120^\circ $ (not drawn here). In this case, two horizontal axes without
any mode-locking are located at $\alpha_3=15^\circ $ and $105^\circ $
instead of $45^\circ $ and $135^\circ $. These axes correspond to
orientations such that the polarization entering the fiber is
elliptical with its high-axis oriented at $45^\circ $ of the
$x$-axis and $y$-axis of the fiber. This is thus similar to previous
cases with $\alpha_4=0^\circ$ or $135^\circ $ and we can understand that no ML
occurs for these two horizontal axes. Note that in this case, two
other axes are observed for $\alpha_3$ near $75^\circ $ and
$165^\circ $. Polarization exiting the plate $n^\circ 3$ is then
circular, which is not modified by the last plate $n^\circ 4$.
Nonlinear polarization rotation is then very difficult to be
obtained, as already mentioned.

We have seen that it is possible to give some physical interpretations concerning
the influence of parameters $\alpha_3$ and $\alpha_4$, located just before the
fiber. Polarization states can then be well understood since
these elements are located just after the polarizer. In contrast, it is
very difficult to interpret the influence of parameters $\alpha_1$ and $\alpha_2$
located at the exit of the fiber. Influence of these parameters depends
indeed strongly on polarization effects induced in the fiber, which are not
directly accessible.
 Experimentally  the role of phase plates
$n^\circ 1$ and $n^\circ 2$ is essential because they allow the adjustment of the polarization state of the incident
electric field at the entrance of  the polarizer, in such a way that the central part of the pulse is transmitted while
the wings are blocked.
However no quantitative description of the influence of the orientation of
phase plates
$n^\circ 1$ and $n^\circ 2$ has been found, due to the high complexity of the nonlinear dynamics.
We but point out their key role.

\begin{figure}[hbt!]
\begin{center}
\includegraphics[width=9cm]{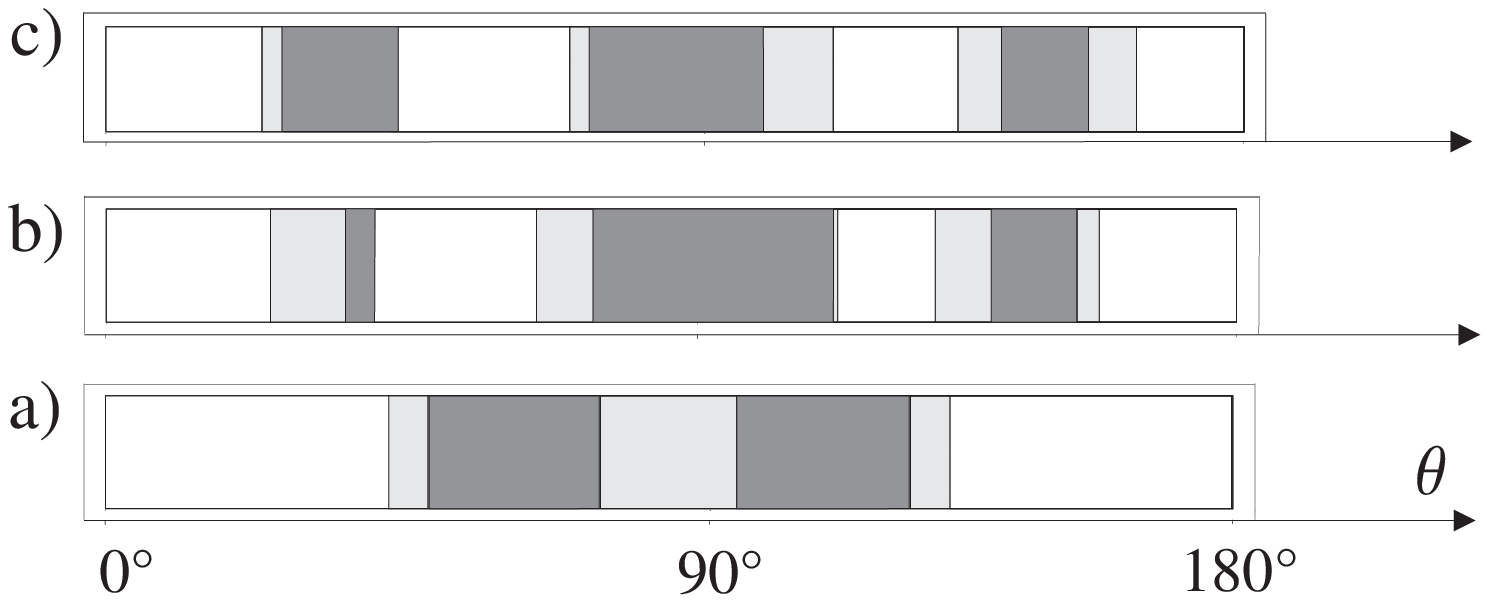}
\caption{\footnotesize Stability of the CW and the mode-locked solutions versus $\theta$ for
$(\alpha_1,\alpha_2,\alpha_3,\alpha_4)=(0^\circ,0^\circ,0^\circ,0^\circ)$ (a),
$(30^\circ,45^\circ,120^\circ,150^\circ)$ (b), $(30^\circ,0^\circ,0^\circ,30^\circ)$
(c). The colors have the same meaning as
 in figure~\ref{carte1}.}
\label{carteta}
\end{center}
\end{figure}

Let us now consider the influence of the orientation of the
polarizer $\theta$ on the operating regimes of the laser for fixed
orientations of the phase plates. Some diagrams are represented in
figure \ref{carteta} for
$(\alpha_1,\alpha_2,\alpha_3,\alpha_4)=(0^\circ,0^\circ,0^\circ,0^\circ)$ (a),
$(30^\circ,45^\circ,120^\circ,150^\circ)$ (b),
and $(30^\circ,0^\circ,0^\circ,30^\circ)$ (c).
 We can note on these figures and also on many
diagrams not reported here that for any values of the orientations
of the phase plates, mode-locking can be achieved by a rotation of
the polarizer.

In summary, although some behaviors can be well interpreted, it is very difficult
to deduce general trends for the mode-locking properties of the laser essentially because
of the large number of variable parameters. However, the model is a very powerful tool
to predict the behavior of the laser.

\section{Conclusion}
In conclusion we have developed a general model for a fiber laser passively mode-locked by nonlinear polarization
rotation. A unidirectional ring cavity containing a polarizer placed between two sets of a halfwave and a
quarterwave plates each has been considered. Starting from two coupled nonlinear
propagation equations for the electric field components
we have derived a unique equation for the field amplitude, which is a complex cubic Ginzburg Landau equation.
The coefficients of the equation depend explicitly on the orientation angles of the polarizer and of the phase plates.
We have thus investigated the stability of both the constant amplitude and the short-pulse solutions as a function
of the angles. Solutions have been found analytically. Although it is difficult to give some general trends, the model
has the advantage to describe a real experiment. Indeed, it includes the linear and nonlinear characteristics of the
doped fiber, two polarization controllers and a polarizer.

\section*{Appendix}
We give hereafter the coefficients of the master equation:
\bq Q=e^{-iKL}\phi_1+e^{iKL}\phi_2,\label{A1}\eq
\bq \phi_1=\left(\chi_1\cos\theta+\chi_2\sin\theta\right)\left(\chi_3\cos\theta+\chi_4\sin\theta\right),\label{A2}\eq
\bq \phi_2=\left(\chi_3^\ast\sin\theta-\chi_4^\ast\cos\theta\right)
\left(\chi_1^\ast\sin\theta-\chi_2^\ast\cos\theta\right),\label{A3}\eq
\bq \chi_1=\frac{-\sqrt 2}2\left[\left(i+\cos(2\alpha_3)\right)\cos(2\alpha_4)
+\sin(2\alpha_3)\sin(2\alpha_4)\right],\label{A4}\eq
\bq \chi_2=\frac{-\sqrt 2}2\left[\left(i-\cos(2\alpha_3)\right)\sin(2\alpha_4)
+\sin(2\alpha_3)\cos(2\alpha_4)\right],\label{A5}\eq
\bq \chi_3=\frac{-\sqrt 2}2\left[\left(i+\cos(2\alpha_2)\right)\cos(2\alpha_1)
+\sin(2\alpha_1)\sin(2\alpha_2)\right],\label{A6}\eq
\bq \chi_4=\frac{-\sqrt 2}2\left[\left(i-\cos(2\alpha_2)\right)\sin(2\alpha_1)
+\cos(2\alpha_1)\sin(2\alpha_2)\right],\label{A7}\eq
and
\bq P=e^{-iKL}\left(\chi_3\cos\theta+\chi_4\sin\theta\right)(\psi_1+\psi_2)+
e^{iKL}\left(\chi_3^\ast\sin\theta-\chi_4^\ast\cos\theta\right)(\psi_3+\psi_4),\label{A8}\eq
with
\bq \psi_1=\gamma B\frac{e^{(2g+4iK)L}-1}{2g+4iK} (\chi_1^\ast\cos\theta+\chi_2^\ast\sin\theta)(\chi_1^\ast\sin\theta-\chi_2^\ast\cos\theta)^2,\label{A9}\eq
\bq \psi_2=\gamma\frac{e^{2gL}-1}{2g}(\chi_1\cos\theta+\chi_2\sin\theta)\biggl[A\left|\chi_1\sin\theta-\chi_2\cos\theta\right|^2
+\left|\chi_1\cos\theta+\chi_2\sin\theta\right|^2\biggl],\label{A10}\eq
\bq \psi_3=\gamma B\frac{e^{(2g-4iK)L}-1}{2g-4iK} (\chi_1\sin\theta-\chi_2\cos\theta)(\chi_1\cos\theta+\chi_2\sin\theta)^2,\label{A11}\eq
\bq \psi_4=\gamma\frac{e^{2gL}-1}{2g}(\chi_1^\ast\sin\theta-\chi_2^\ast\cos\theta)\biggl[A\left|\chi_1\cos\theta+\chi_2\sin\theta\right|^2
+\left|\chi_1\sin\theta-\chi_2\cos\theta\right|^2\biggl].\label{A12}\eq

\end{document}